\documentclass[aps,pra,epsfigure,showpacs,twocolumn]{revtex4}

\usepackage{dcolumn}

\usepackage{bm}

\usepackage{graphicx}

\usepackage{amsmath}

\usepackage{latexsym}

\usepackage{amsfonts}

\usepackage{amssymb}

\usepackage{array}

\usepackage{epsfig}

\newcommand{\ket}[1]{\left\vert#1\right\rangle}

\newcommand{\bra}[1]{\left\langle#1\right\vert}

\begin{document}

\title{Towards variance-matrix characterization of complementarity relations 

in a continuous variable system}

\author{S. Springer and M. Paternostro}

\affiliation{School of Mathematics and Physics, Queen's University, Belfast BT7 1NN, United Kingdom}

\date{\today}

\begin{abstract}

We discuss complementarity relations in a bipartite continuous variable system. Building up from the work done on discrete $d$-dimensional systems, we prove that for symmetric two-mode states, quantum complementarity relations can be put in a simple relation with the elements of the variance matrix. When this condition is not satisfied, such a connection becomes non-trivial. Our investigation is the first step towards an operative characterization of the complementarity in a scenario that has not been investigated so far. 

\end{abstract}


\maketitle

\section{Complementarity}

Bohr's principle of complementarity~\cite{BohrComp} is a cornerstone of the modern understanding of the physical world at the quantum level. It states that a quantum system may exhibit equally real yet mutually exclusive properties--it requires seperate experiments to examine different features of a system. Certain properties cannot be measured simultaneously even though both measurements are individually possible and are in fact necessary for a full description of the object. This principle forms the foundation of the Copenhagen interpretation of quantum mechanics and states that a quantum object, such as an electron, does not have an objective reality; instead all descriptions of properties of such an object must be given in terms of experiments made on the object. We do not `reveal' the value of properties in experiments but rather assign them by the act of measurement. 

The original \emph{orthodox} complementarity principle of Bohr states that wave and particle attributes may not be simultaneously displayed by a quantum system. This stringent stipulation was later relaxed to permit the observation of complementary observables within the same experimental set up, although retaining a restriction on the degree to which each could be observed in the same experiment. Wooters and Zurek~\cite{WootersZurek} were the first to consider an interferometric setting where particle-like {\it welcher weg} information was obtained whilst preserving the wave-like interference pattern, although an increase in the precision with which one of the two behaviors is observed results in a corresponding decrease in the precision of the other. A similar observation concerning neutron interferometry was made by Greenberger and Yasin a decade later~\cite{GreenbergerYasin}. A quantative theoretical basis for this effect has been given independently by Jaeger \emph{et al}~\cite{JaegerEtAl} and Englert~\cite{Englert}. 

More recently, Jakob and Bergou~\cite{JakobBergou1} derived a generalised complementarity relation for pure bipartite systems of two qubits involving local properties (that describe the single-partite nature of the system) and a measure of the non-local properties that describe its bipartite nature. A proposal for a quantum-non-demolition test of bipartite complementarity relations has been put forward in Ref.~\cite{davidovich}. Here, we study bipartite complementarity in a continuous variable (CV) system by taking the infinite-dimensional limit of the relation designed in~\cite{JakobBergou2} for $d$-level systems. Our aims are firm on a pragmatic ground: we are interested in verifying quantum complementarity between single-partite and multi-partite properties of a two-mode CV system and in identifying a clear connection with experimentally-accessible quantities characterizing the state at hand. We achieve this result by using the powerful tool of variance-matrix formalism of two-mode Gaussian states~\cite{barnettradmore}. This class encompasses CV states whose characteristic function is Gaussian in the phase space and includes many relevant experimentally available CV states. We show that for two-mode states with specific chracteristics of symmetry, any entry in the bipartite complementarity relation can be expressed in terms of the elements of the variance matrix of the system and, as such, can be fully experimentally determined by means of homodyne measurements, for instance~\cite{myungmunro}.

This paper is organized as follows. In Sec.~\ref{complementarity} we briefly discuss the details of bipartite complementarity relations and their extension to $d$-level systems. This serves as a basis for the discussion in Sec.~\ref{center}, where we quantitatively address the case of pure bipartite CV systems, after a concise introduction to the formal tools used throughout our investigation. We also discuss the possibility of extending our findings to non-symmetrical cases. Finally, in Sec.~\ref{conclusions} we summarize our results.

\section{Bipartite complementarity relation}

\label{complementarity}

In order to fix the ideas and pave the way to our analysis, we first briefly discuss the work in Ref.~\cite{JakobBergou1}, where a three-term complementarity relation for pure bipartite qubit systems has been introduced, involving local and non-local properties of a system.

We consider a general, pure state of two qubits (indicated as $a$ and $b$): $\ket{\psi}_{ab} =  a_0\ket{00}_{ab} + a_1\ket{01}_{ab} + a_2\ket{10}_{ab} + a_3\ket{11}_{ab}$ (assumed to be normalized, so that $\sum^3_{i=0}|a_i|^2=1$), where $\{\ket{0}_{i},\ket{1}_{i}\}$ stand for the two states of a qubit $i=a,b$. We then introduce the {\it visibility} relative to qubit $i$

\begin{equation}
\mathcal{V}_i = 2|\bra{\psi} \hat{\sigma}_+^i \ket{\psi}| \qquad \textrm{with} \qquad \hat{\sigma}_+^i = \begin{pmatrix}0&1\\0&0\end{pmatrix},
\end{equation}
the corresponding {\it predictability}  

\begin{equation}
\mathcal{P}_i = |\bra{\psi}  \hat{\sigma}_z^i \ket{\psi}| \qquad \textrm{with} \qquad \hat{\sigma}_z^i =\begin{pmatrix}1&0\\0&-1\end{pmatrix}
\end{equation} 
and, finally, we consider the pure-state form of {\it concurrence}~\cite{concurrence}

\begin{equation}
\label{conc}
\mathcal{C}=|\bra{\psi^*} (\hat{\sigma}^a_y \otimes\hat{\sigma}^b_y) \ket{\psi}|\equiv\sqrt{2(1-\text{Tr}_i[(\varrho^{i})^2])}
\end{equation}
with $\hat{\sigma}^i_y=i(\hat{\sigma}^{i\dag}_+-\hat{\sigma}^i_+)$, $\ket{\psi^*} =  a^*_0\ket{00}_{ab} + a^*_1\ket{01}_{ab} + a^*_2\ket{10}_{ab} + a^*_3\ket{11}_{ab}$ and $\varrho^i=\text{Tr}_{i'\neq{i}}\ket{\psi}_{ab}\!\bra{\psi}$ ($i,i'=a,b$). The first two quantities give a measure of single-particle behavior in that they account for single-particle coherence and welcher-weg information ({\it i.e.} single-particle relative population), respectively. On the other hand, concurrence accounts for genuine bipartite properties of the state under scrutiny. With these definitions, a three-entry complementarity relation can be formulated as~\cite{JakobBergou1,davidovich}
\begin{equation}
(\mathcal{V}_i^2 + \mathcal{P}_i^2) + \mathcal{C}^2  = 1 \label{trialityrelation} 
\end{equation}
Visibility and predictability in Eq.~(\ref{trialityrelation}) have been grouped together in order to highlight their inter-convertibility under local unitary transformation and the fact that they account for genuine single-partite character. 

With this understanding, it is appropriate to think about a collective single-particle quantity, defined as ${\cal S}_i^2 = {\cal V}_i^2 +{\cal P}_i^2$, which is globally invariant under local unitary operations. We may then express the complementarity relation as a more traditional ``duality'' formula
\begin{equation}
{\cal C}^2 + {\cal S}_i^2 =1.
\end{equation}
It is worth stressing, on the other hand, that a grouping process putting together concurrence and predictability, so as to form a new entity called {\it distinguishability}, allows us to put bipartite complementarity relations of the form reported here in correspondence with Englert's complemetarity relation~\cite{Englert}, which is only apparently a single-partite relation~\cite{correlations}. 

Key to the aims of our work is the observation that Eq.~(\ref{trialityrelation}) can be straightforwardly extended to pure states of bipartite $d$-dimensional systems ($d\in\mathbb{Z}$ is assumed to be the same for both the subsystems). Hioe and Eberly~\cite{HioeEberly} suggested how to write the components of a $d$-dimensional state's Bloch vector as the expectation values of a complete set of observables $ \{ \hat{u}\}, \{\hat{v}\}$ and $\{\hat{w}\}$ defined, for subsystem $i$, as
\begin{equation}
\begin{split}
\hat{u}^i_{jk} &= | k\rangle_i\langle j|+| j\rangle_i\langle k|,\\
\hat{v}^i_{jk} &= i(| k\rangle_i\langle j|-| j\rangle_i\langle k|),\\
\hat {w}^i_l &= \sqrt{{2}/{l(l+1)}}[ \displaystyle \sum_{j=1}^{l}  (| j\rangle_i\langle j| - | l+1\rangle_i\langle l+1|)].
\end{split}
\end{equation}
where the indices are constrained by $1 \le j < k \le d,~1 \le l \le d-1$. With this formalism and by using the length of the Bloch vector as a measure of the information content of the system, it is possible to express the key quantities of a dual complementarity relation~\cite{JakobBergou2}.

Specifically, the visibility is written in terms of $\hat{u}_{jk}$'s and $\hat{v}_{jk}$'s, which are associated with state correlations, as

\begin{equation}
\label{visibilita}
\mathcal{V}^2_i =  \displaystyle \sum_{j,k; j < k} (|\langle \hat{u}^i_{jk}\rangle|^2 + |\langle \hat{v}^i_{jk} \rangle|^2) = 2\!\sum_{j,k; j \neq k} |\varrho^i_{jk}|^2,
\end{equation}
where $\varrho^i_{jk}={}_i\!\bra{j}\varrho^i\ket{k}_{i}$. The predictibility is linked to the operators which account for population statistics, namely the $\hat{w}^i_l$, as

\begin{equation}
\label{predicibilita}
\mathcal{P}^2_i =  \displaystyle \sum_{l=1}^{d-1} |\langle \hat{w}^i_l \rangle|^2 = 2[\sum_{j=1}^n (\varrho^i_{jj})^2 - \frac{1}{d}]
\end{equation}
Thus the single-partite nature of the system is given by $\mathcal{S}^2_i = \mathcal{P} ^2_i + \mathcal{V}^2_i= 2~\text{Tr}[(\varrho^i)^2]-\frac{2}{d}$.

This last quantity is top-bounded by $\frac{2(d-1)}{d}$ (due to $\text{Tr}[(\varrho^i)^2] \le 1$). A term accounting for the non-local aspects of the quantum system should be included. This can be done by taking the natural generalization of the spin-flip operation involved in the definition of concurrence, as given in Eq.~(\ref{conc}). Such a tool is provided by the universal-inverter operator and the associated generalization of our entanglement measure is given by the I-concurrence introduced in~\cite{iconc} and defined as $\mathcal{C}^2_I = 2- 2\text{Tr}[(\varrho^i)^2]$. We finally get~\cite{JakobBergou2}

\begin{equation}
\label{startingpoint}
\mathcal{P}^2_i + \mathcal{V}^2_i + \mathcal{C}^2_I \le \frac{2(d -1)}{d}.
\end{equation}
This complementarity relation for finite $d$-dimensional subsystem is the starting point of our study, as described in the next Section.

\section{Complementarity relation for fully symmetric CV systems}

\label{center}

Our study is restricted to two-mode Gaussian states of CV systems, which are efficiently described in terms of the associated variance matrix (VM). For completeness, we now provide the tools used in our investigation and then assess the connection between VM and complementarity.

\subsection{Introduction to CV formalism}

Let two CV systems be described by the respective phase-space quadrature operators $\hat{x}_i$ and $\hat{p}_{i}$ (with $[\hat{x}_i,\hat{p}_i]=i$) and let us assume they are prepared in a Gaussian state. We can build up the VM ${\mathbf V}$ of elements~\cite{AdessoIlluminati,MyungReview}

\begin{equation}
V_{jk} =\langle \{\hat{q}_j,\hat{q}_k \} \rangle - 2 \langle\hat{q}_j\rangle \langle\hat{q}_k\rangle\label{Vijdef}~~~~(i,j=1,..,4),
\end{equation}
where $\hat{\mathbf q}=(\hat{x}_a,\hat{p}_a,\hat{x}_b,\hat{p}_b)$. The Wigner function of a Gaussian state, and thus the corresponding state, is completely determined by the assignment of $\mathbf{V}$~\cite{barnettradmore}, which therefore contains full information on the local and joint properties of a multi-mode system. For the bipartite case, a VM can always be written as

\begin{equation}
{\bf V}=
\begin{pmatrix}
{\bf A}&{\bf C}\\
{\bf C}^T&{\bf B}
\end{pmatrix},
\end{equation}
where ${\bf A}$ and ${\bf B}$ are $2\times{2}$ matrices accounting for the local properties of systems $a$ and $b$, respectively, while ${\bf C}$ characterizes any inter-system correlations. The set of operations that preserve the Gaussian nature of a state in the phase-space representation are the {\it symplectic operations}---analogous to unitary operators acting on the Hilbert space. A symplectic transformation, $S$, is one such that $S^T \Omega S= \Omega$, with ${\bm \Omega}=\bigoplus_{j=a}^bi\hat{\bm\sigma}^j_y$ the so-called symplectic matrix. The set of all joint symplectic operations on a two-mode state is labelled $Sp (4,\mathbb{R})$ while local symplectics are represented by $Sp(2, \mathbb{R})^{\oplus2}$. 

The positive-semidefiniteness of density matrices is translated, in the VM formalism, into the Heisenberg-Robertson condition

\begin{eqnarray}
{\mathbf V} + i \bm{\Omega} \ge 0.
\end{eqnarray}
It should be clear that the key advantage of the VM formalism is the finiteness of ${\bf V}$, despite the infinite dimension of the Hilbert spaces of the modes under investigation. Any property (such as purity and entanglement) of the state can be expressed by relying on the elements of the associated VM, which is characterized by important properties of invariance with respect to symplectic operations. For instance, it is easy to recognize the symplectic invariance of $\text{det}{\mathbf V}$ and $\Delta=\text{det}{\mathbf A}+\text{det}{\mathbf B}+2~\text{det}{\mathbf C}$~\cite{AdessoIlluminati}. This is particularly relevant as convenient formulae for entanglement measures such as {negativity}~\cite{nega} can be given in terms of {\it symplectic eigenvalues}. These, which are formally defined as othogonal eigenvalues of $i{\mathbf \Omega}{\mathbf{V}}$, are explicitly given by $\nu_{\pm} = \frac{1}{\sqrt{2}} \sqrt{ \Delta \pm \sqrt{\Delta^2 - 4~\text{det}(\mathbf{V})}}$. In terms of symplectic eigenvalues, the separability criterion based on positivity of partial transposition~\cite{peres} is rephrased as $|\tilde\nu'|\ge1$, where $\tilde\nu'$ is the smallest symplectic eigenvalue of $i{\mathbf \Omega}{(\openone\oplus{\bm\sigma}_z)V(\openone\oplus{\bm\sigma}_z)}$~\cite{AdessoIlluminati}.

\subsection{VM-formulation of bipartite complementarity relation} 

We now focus on the central aim of our study, {\it i.e.} the formulation of an operative connection between the VM entries and the bipartite complementarity relation. Starting from Eq.~(\ref{startingpoint}), we will consider a truncated pure two-mode CV state and explicitly construct the corresponding VM for any dimension of the associated Hilbert space. Such state should respect the $d$-dimensional bipartite complementarity relation of Eq.~(\ref{startingpoint}). The transition to the infinite-dimensional case will be performed by smoothly considering the limit $d\rightarrow{\infty}$.

As the prototype of our state and being motivated by reasons of practicality of its generation, we consider a truncated state 

\begin{equation}
\label{trunc}
| \psi(t) \rangle_{ab}=\mathcal{N}(\xi)\sum_{n=0}^{t} \xi^{n} |n, n \rangle_{ab}
\end{equation}
where $|n\rangle_i$ is an $n$-photon Fock state of mode $i=a,b$ and the shorthand $\mathcal{N}(\xi) = \sqrt{\frac{1-\xi^2}{1-\xi^{2t +2}}}$ with $\xi=\tanh{r}$ has been used. In Eq.~(\ref{trunc}), $t$ is a cut-off parameter that top-bounds the dimension of the Hilbert space of modes $a$ and $b$. The local and nonlocal properties of $\ket{\psi(t)}_{ab}$ depend on $r$ and, in general, on the cut-off parameter. For an infinite cut-off, $\ket{\psi(\infty)}_{ab}$ becomes the standard two-mode squeezed vacuum state $\ket{\text{TMSS}}_{ab}= (\cosh{r})^{-1}\sum_{n=0}^{\infty} \xi^n |n,n \rangle_{ab}$ of squeezing parameter $r$, which thus provides an operative interpretation for the parameter $r$ entering Eq.~(\ref{trunc}). 

\begin{figure}[b]
\begin{center}
\includegraphics[width=0.45\textwidth]{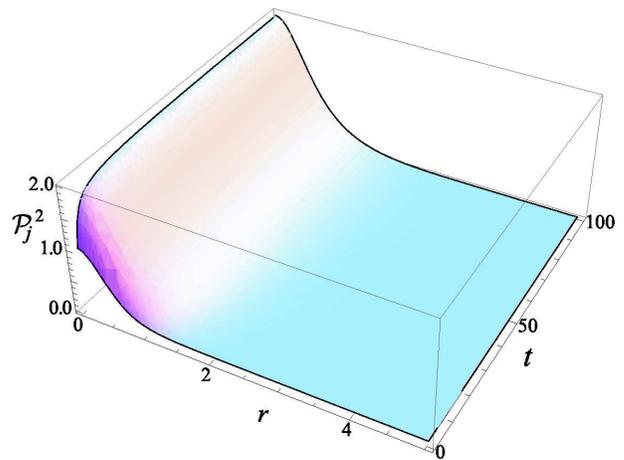}
\caption{Predictability for a truncated two-mode squeezed vacuum against the squeezing factor $r$ and the cut-off value $t$.}
\label{predictability}
\end{center}
\end{figure}

\begin{figure}[b]
\begin{center}
\includegraphics[width=0.45\textwidth]{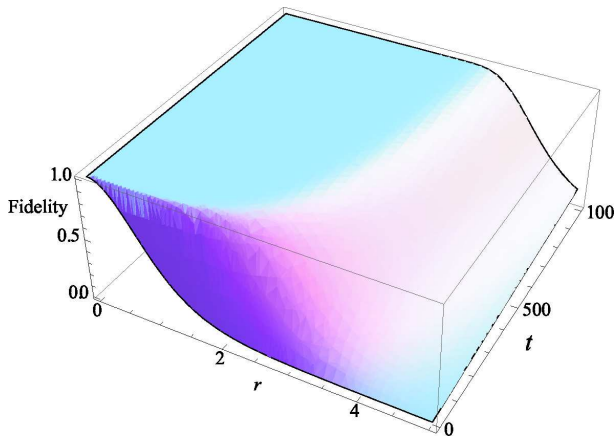}
\caption{Fidelity between the truncated state $\ket{\psi(t)}_{ab}$ and a two-mode squeezed state, plotted against squeezing $r$ and cut-off $t$. For a set value of squeezing, it is enough to increase $t$ in order to achieve perfect state fidelity.}
\label{fidelity}
\end{center}
\end{figure}

The evaluation of the components of the bipartite complementarity relation requires the density matrix of subsystem $j$ 

\begin{equation}
\varrho_j= \mathcal{N}^2(\xi)\sum_{n=0}^{t} \xi^{2n} |n\rangle_j\langle n |~~~~(j=a,b)
\end{equation}
with $\varrho_a=\varrho_b$ given the symmetry of the bipartite state. Since such a reduced state is diagonal in the Fock-basis, the coherences will be zero, which immediately sets the visibility~(\ref{visibilita}) equal to zero. On the other hand, the predictibility Eq.~(\ref{predicibilita}) is straightforwardly evaluated to be

\begin{equation}
\label{predicibilita2}
\begin{split}
\mathcal{P}^2_j(\xi,t)&=2[{\cal N}^4(\xi)\sum^t_{n=0}\xi^{4n}-\frac{1}{t+1}]\\
&= 2\left[\mathcal{N}^4(\xi){\frac{1-\xi^{4t +4}}{1-\xi^4}} - \frac{1}{t+1}  \right],
\end{split}
\end{equation}
where the last term accounts for the $(t+1)$-dimensionality of the Hilbert space of each mode. 

\begin{figure}[b]
\begin{center}
\includegraphics[width=0.45\textwidth]{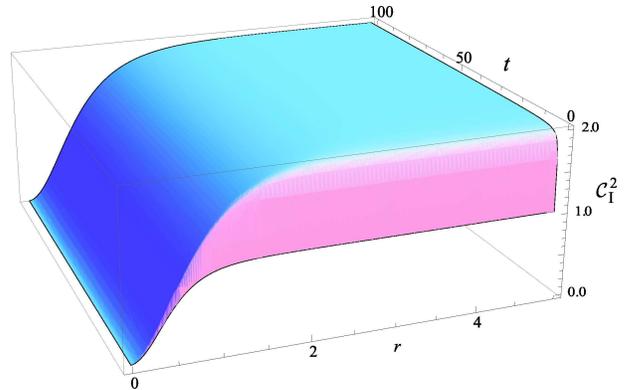}
\caption{I-concurrence for a truncated two-mode squeezed state against the squeezing factor $r$ and the cut-off value $t$.}
\label{concurrence}
\end{center}
\end{figure}

The behavior of predictability is shown in Fig.~\ref{predictability} versus the squeezing parameter $r$ and the value of the cut-off. While the trend against $t$ is almost uniform and is quickly stabilized to stationary values, ${\cal P}^2_j(\xi,t)$ is a rapidly decreasing function of $r$, starting from $2$ for $r=0$ and becoming null soon after $r\sim{1.5}$, uniformly with respect to $t$. Physically, this implies the disappearance of {\it welcher weg-like} information associated with a bias in the population of the Fock-states entering Eq.~(\ref{trunc}). These considerations obviously apply to the standard two-mode squeezed vacuum in the region of parameters where $\ket{\psi(t)}_{ab}$ represents a good approximation of $\ket{\text{TMSS}}_{ab}$. This occurs as illustrated in Fig.~\ref{fidelity}, where the state fidelity~\cite{NC}

\begin{equation}
|\bra{\psi(t)}\text{TMSS}\rangle|^2=\left(\frac{\cal N}{\cosh{r}}\sum^t_{n=0}\xi^{2n}\right)^2=\frac{1}{{\cal N}^2\cosh^2{r}}
\end{equation}
shows that, as intuitively expected, an increase in $r$ requires a larger cut-off in order to gain sufficient closeness of the two states. In a wide area where the state fidelity is, for all practical purposes, equal to $1$, we have that ${\cal P}^2_j(\xi,t)\simeq0$. As the single-partite manifestations of $\ket{\psi(t)}_{ab}$ are all ascribed to the predictability, we now want to connect this result to the expected gain of strength in the bipartite character of the state at hand. To this end, we refer to the I-concurrence, which is found to give

\begin{equation}
\label{concorrenza}
\mathcal{C}^2_{I}(\xi,t)=2\left(1- \mathcal{N}^4(\xi) \frac{1-\xi^{4t +4}}{1-\xi^4}\right)
\end{equation}

and is plotted in Fig.~\ref{concurrence}. Here the I-concurrence is asymptotically bounded by $2$ as $r$ and $t$ grow, in agreement with the bound expected for an infinite-dimensional systems. This occurs quickly in $r$ and persists in those regions of the relevant parameters where the state fidelity is non-ideal. In particular, by inspection of Eqs.~(\ref{predicibilita}) and (\ref{concorrenza}), we get that, for any value of $r\neq{0}$, the bound is

\begin{equation}
{\cal P}^2_j(\xi,t)+{\cal C}^2_{I}(\xi,t)=\frac{2t}{t+1}\mathop{\longrightarrow}\limits_{t\to\infty}{2}.
\end{equation}
From Fig.~\ref{fidelity}, we know that at a fixed $r$ it is enough to take a sufficiently large value of the cut-off $t$ in order to have an ideal state fidelity. This implies that genuine two-mode squeezed vacuum states of finite squeezing saturate the bound imposed on a bipartite complementarity relation.

Having examined in detail the form that each entry in Eq.~(\ref{startingpoint}) has, we now turn to the properties of the VM associated with $\ket{\psi(t)}_{ab}$. Regardless of the value of $t$, it is immediately recognizable that the expectation values of single-quadrature operators of Eq.~(\ref{trunc}) are identically zero, which simplifies the VM elements to $V_{ij} =\langle \{ \hat{q}_j,\hat{q}_k \} \rangle$. It is matter of straightforward calculations to show that 

\begin{equation}
\label{covariance}
  {\mathbf V}_{\psi}= 
\begin{pmatrix}

        V_{11} & 0 & V_{13} & 0 \\

      0 & V_{11} & 0 &  -V_{13}\\

      V_{13} & 0 & V_{11} & 0  \\

      0 &  -V_{13}  & 0 & V_{11} 

\end{pmatrix}
\end{equation}
with the elements being explicitly given by

\begin{equation}
\label{elements}
\begin{split}
V_{11}& =
\mathcal{N}^2(\xi) \displaystyle \sum_{n=0}^t \xi^{2n}  (2n+1)=  {\cal N}^2(\xi)\partial_{\xi}\left(\frac{\xi}{{\cal N}^2(\xi)}\right)\\ 
&=\frac{1+\xi^2-(3+2t)\xi^{2t+2}+(1+2t)\xi^{2t+4}}{(1-\xi^2)(1-\xi^{2t+2})},\\
V_{13}
&=2 \mathcal{N}^2(\xi) \sum_{n=0}^{t-1} \xi^{2n+1}(n+1)\\
&=2\xi\frac{1-\xi^{2t}(1-\xi^2)-\xi^{2t}[t+(1-t)\xi^2]}{(1-\xi^2)(1-\xi^{2t+2})}.\\ 
\end{split}    
\end{equation}
It can be proved that $V_{11}>1, \forall{r,t}$. In the limit of $t\rightarrow\infty$, $V_{11}\rightarrow\cosh(2r)$ and  $V_{13}\rightarrow\sinh(2r)$ and we gain back the VM of a proper two-mode squeezed vacuum state

\begin{equation}
{\mathbf V}_{\text{TMSS}}=
\begin{pmatrix}
\cosh(2r)\openone&\sinh(2r){\bm\sigma_{z}}\\
\sinh(2r){\bm\sigma_{z}}&\cosh(2r)\openone
\end{pmatrix}
\end{equation}

\begin{figure}[b]
\begin{center}
\includegraphics[width=0.45\textwidth]{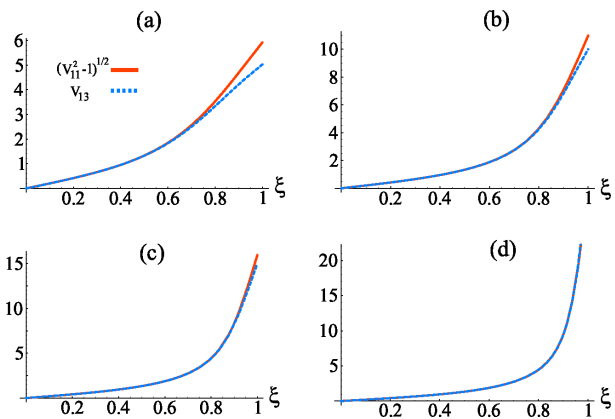}
\caption{Discrepancy between $V_{13}$ (dotted line) as determined by the second of Eqs.~(\ref{elements}) and the value $\sqrt{V^2_{11}-1}$ (solid line) determined by imposing perfect purity of an infinite-dimensional Gaussian state. The horizontal axis show $\xi=\tanh{r}\in[0,1]$ for $r\in[0,\infty)$. From panel {\bf (a)} to {\bf (d)}, $t$ goes from $5$ to $20$ with incremental steps of $5$.}
\label{andamento}
\end{center}
\end{figure}
However, it is important to stress that for a finite value of $t$, $\ket{\psi(t)}_{ab}$ is not a Gaussian state. As a consequence, this prevents the use of relations connecting the value of $V_{13}$ to that of $V_{11}$ as based, for instance, on purity considerations~\cite{AdessoIlluminati}. In fact, would Eq.~(\ref{covariance}) have been the VM of a pure Gaussian state, the purity of the state could have been written as 

\begin{equation}
\label{purity}
\frac{1}{\sqrt{\text{det}{\mathbf V}_{\psi}}}=\frac{1}{V^2_{11}-V^2_{13}}=1,
\end{equation}
thus setting $V_{13}=\sqrt{V^2_{11}-1}$. Such a condition holds only approximately for small $r$ and requires large values of $t$ in order to be accurate up to $\xi\simeq{1}$, as it is illustrated in Fig.~\ref{andamento}. Clearly, for $t\rightarrow\infty$ we recover the Gaussian case and the validity of the constraint imposed by purity considerations holds rigorously.  

Nevertheless, a particularly simple relation between the only two non-zero entries of the VM at hand is found to be

\begin{equation}
\xi=\frac{V_{11}-1}{{V}_{13}},
\end{equation}
as it is easy to verify using Eqs.~(\ref{elements}). This allows us to explicitly connect both the predictability and the I-concurrence for $\ket{\psi(t)}_{ab}$ to the elements of the VM corresponding to the state. Indeed, 

\begin{equation}
\label{relation}
{\cal N}^4(\xi)\frac{1-\xi^{4t+4}}{1-\xi^4}\simeq\frac{V^2_{13}-(V_{11}-1)^2}{V^2_{13}+(V_{11}-1)^2},
\end{equation}

which is valid for $t\gg{1}$~\cite{commento}. 

Eq.~(\ref{relation}) can now be replaced in the expressions for predictability and I-concurrence, Eqs.~(\ref{predicibilita2}) and (\ref{concorrenza}), so as to relate the bipartite complementarity relation to the elements of the VM of a fully-symmetric state having a finite degree of squeezing. It is interesting to look at the limit $t\rightarrow\infty$, so that the purity-based constraint $V^2_{13}=V^2_{11}-1$ can be taken, leading to 

\begin{equation}
\label{finalsymmetric}
{\cal C}^2_I=2\left(1-\frac{1}{V_{11}}\right),
\end{equation}
in agreement with the fact that I-concurrence is proportional to the linear entropy of entanglement~\cite{giampaolo}. Therefore, we recognize that for a pure two-mode squeezed state, bipartite complementarity can be probed simply by experimentally determining the variance of one of the quadratures of its modes. This will immediately give us a quantitative estimate of the balance between predictability and entanglement in the state. The fact that the relation of bipartite complementarity is fully determined, in this case, by just one parameter should not come as a surprise given that the VM of a pure bipartite state is fully determined by $V_{11}$.

The result above can be directly extended to other situations. In fact, any pure entangled two-mode state can be reduced, via single-mode squeezing and rotations, to the structure that has been studied in this paper, where visibility is made to disappear and complementarity is entirely ascribed to the interplay between predictability and entanglement. This can be seen as a manifestation of the inter-convertibility of visibility and predictability, as stated in Sec. \ref{complementarity}. In fact, the class of operations required in order to get a VM of the form of ${\mathbf V}_\psi$ comprises only local operations, which can just ``re-arrange" the distribution of single-body properties of a state, rather than the balance between single-partite and bipartite ones.

As an illustrative example, we can consider the case of another experimentally relevant two-mode state generated by superimposing a single-mode squeezed state to the vacuum at a $50:50$ beam splitter~\cite{barnettradmore}. The resulting state can be written as~\cite{BuzekVidiella}

\begin{equation}
\label{single}
\ket{\xi}_{ab}=\frac{1}{\sqrt{2\pi\sinh{r}}}\int{d}\alpha{e}^{-\frac{1-\xi}{2\xi}\alpha^2}|{\alpha}/{\sqrt 2},{\alpha}/{\sqrt 2}\rangle_{ab}
\end{equation}
with $\alpha\in{\mathbb R}$. Differently from the case of a TMSS treated above, the reduced density matrix of one of the modes is non-diagonal in the Fock-basis. Indeed, we have that

\begin{equation}
\begin{split}
\varrho^{j}&=\frac{1}{{2\pi\sinh{r}}}\sum^{\infty}_{n,m=0}\frac{{\cal I}_{nm}}{2^{\frac{n+m}{2}}\sqrt{n!m!}}\ket{n}_{j}\!\bra{m}
\end{split}
\end{equation}
with ${\cal I}_{nm}=\int{d}\beta{e}^{-\frac{1+\xi}{2\xi}(\beta^2)}\beta^m\Sigma_n(\beta)$ and the formula~\cite{rykiz}

\begin{equation}
\Sigma_{n}(\beta)\!=\!\int{d}\alpha{e}^{-\frac{1+\xi}{2\xi}\alpha^2+\alpha\beta}\alpha^n
\!=\!\frac{4\xi^{3/2}\sqrt{2\pi}}{2^{n}(1+\xi)^{3/2}}\partial^{n-1}_{q^{n-1}}qe^{\frac{2\xi{q}^2}{2(1+\xi)}}.
\end{equation}
where $q=\beta/2$. Numerically, it can be checked that, in general, $\varrho^{j}_{nm}\neq{0}$, thus indicating a non-zero visibility associated with the state. However, its numerical nature makes the connection between bipartite complementarity and VM elements rather impractical. However, one can resort to the inter-convertibility of single-partite properties. More specifically, the VM associated with $\ket{\xi}_{ab}$ is easily found to be

\begin{equation}
{\mathbf{V}}_{\xi}=
\begin{pmatrix}

e^r\cosh{r}&0&e^r\sinh{r}&0\\

0&e^{-r}\cosh{r}&0&-e^{-r}\sinh{r}\\

e^{r}\sinh{r}&0&e^{r}\cosh{r}&0\\

0&-e^{-r}\sinh{r}&0&e^{-r}\cosh{r}

\end{pmatrix}
\end{equation}
which reveals a clear asymmetry between the $\hat{x}$ and $\hat{q}$ quadrature of each mode. It is simple to recognize that local anti-squeezing represented by the symplectic operation $e^{-\frac{r}{2}{\bm \sigma}_z}\oplus{e^{-\frac{r}{2}{\bm \sigma}_z}}$ puts ${\mathbf V}_{\xi}$ into a form corresponding to a TMSS of squeezing parameter $r/2$, which can then be treated following precisely the same approach as described above. 

\section{Conclusions}

\label{conclusions}

We have discussed the relation between the VM of a bipartite CV state and a complementarity relation that contrasts local and non-local properties. We have revealed that symmetry in the VM allows the formulation of a simple connection between quadrature variances and predictability/entanglement. The pragmatic relevance of such a connection should be rather clear. The simplicity of our findings make the assessment of the local-against-global properties of a state experimentally feasible by means of homodyne detection, as suggested in Refs.~\cite{myungmunro,Leonhardt}. This investigation may prompt an experimental test of complementarity in CV systems.

\acknowledgments

We thank Prof. M. S. Kim for helpful discussions and encouragement regarding this work.
SS thanks Prof. J. Bergou for having kindly sent him a preprint of Ref.~\cite{JakobBergou2}. We acknowledge financial support from DEL and UK EPSRC. MP acknowledges the Bridging Fund from Queen's University Belfast.


\begin{thebibliography}{25}
\bibitem{commento} It is straightforward to relax the condition of large $t$ in Eq.~(\ref{relation}), whose right-hand side is generalized to the much more involved relation 
\begin{equation}
\frac{[V^{2t+2}_{13}+(V_{11}-1)^{2t+2}][V^2_{13}-(V_{11}-1)^2]}{[V^{2t+2}_{13}-(V_{11}-1)^{2t+2}][V^2_{13}+(V_{11}-1)^2]}.
\end{equation}

This expression holds for any $r$ and $t$.
\end{thebibliography}
\end{document}